\documentclass[twocolumn,10pt,journal,comsoc]{IEEEtran}
\usepackage{cite}
\usepackage{amsmath,amssymb,amsfonts}
\usepackage{algorithmic}
\usepackage{graphicx}
\usepackage{textcomp}
\usepackage{xcolor}
\usepackage{epstopdf}
\def\BibTeX{{\rm B\kern-.05em{\sc i\kern-.025em b}\kern-.08em
    T\kern-.1667em\lower.7ex\hbox{E}\kern-.125emX}}

\begin{document}
\title{Task Admission Control and Boundary Analysis of Cognitive Cloud Data Centers
}

\author{\IEEEauthorblockN{Wenlong Ni\IEEEauthorrefmark{1},
Yuhong Zhang\IEEEauthorrefmark{2}, and
%Montgomery Scott\IEEEauthorrefmark{3}, and
Wei Li\IEEEauthorrefmark{2},~\IEEEmembership{Senior Member,~IEEE}}

\IEEEauthorblockA{\IEEEauthorrefmark{1}99 ZiYang Ave, JiangXi Normal University NanChang, CHINA 330022}

\IEEEauthorblockA{\IEEEauthorrefmark{2}3100 Cleburne St, Texas Southern University, Houston, USA TX 77004}
\thanks{This work was supported in part by US National Science Foundation under Grant No. CNS1827940 and JiangXi Education Department under Grant No. GJJ191688.}
}

\maketitle

\begin{abstract}
A novel cloud data center (DC) model is studied here with cognitive capabilities for real-time (or online) flow compared to the batch tasks. Here, a DC can determine the cost of using resources and an online user or the user with batch tasks may decide whether or not to pay for getting the services. The online service tasks have a higher priority in getting the service over batch tasks. Both types of tasks need a certain number of virtual machines (VM). By targeting on the maximization of total discounted reward, an optimal policy for admitting task tasks is finally verified to be a state-related control limit policy. Next, a lower and an upper bound for such an optimal policy are derived, respectively, for the estimation and utilization in reality. Finally, a comprehensive set of experiments on the various cases to validate this proposed model and the solution is conducted. As a demonstration, the machine learning method is adopted to show how to obtain the optimal values by using a feed-forward neural network model. The results achieved in this paper will be expectedly utilized in various cloud data centers with cognitive characteristics in an economically optimal strategy.

\end{abstract}

%\fi
%\newpage
\section{Introduction}
With the rapid development of Internet and computing technology, more and more data are continually being produced from all over the world. Nowadays, cloud computing has become fundamental for IT operations worldwide, replacing traditional business models. Enterprises can now access the vast majority of software and services online through a visualized environment, avoiding the need for expensive investments in IT infrastructure. On the other hand, challenges in cloud computing need to be addressed such as security, energy efficiency and automated service provisioning. According to various studies, electricity use by DCs has increased significantly due largely to explosive growth in both the number and density of data centers (DCs)\cite{osti_1372902}. To serve the vast and rapidly growing needs of online businesses, infrastructure providers often end up over provisioning their resources (e.g., number of physical servers) in the DC. Energy efficiency in DCs is a goal of fundamental importance. Various energy-aware scheduling approaches have been proposed to save power and energy consumptions in DCs by minimizing the number of active physical servers hosting the active virtual machines (VMs); VMs also can be migrated between different hosts if necessary\cite{closer17,7572038,8267099}.

There are basically two parties in the cloud computing paradigm: the cloud service providers (CSPs) and the clients, who each have their own roles in providing and using the computing resources. While enjoying the convenient on-demand access to computing resources or services, the clients need to pay for this access. CSPs can make a profit by charging their clients for the services provided. Clients can avoid the costs associated with 'inhouse' provisioning of computing resources and at the same time have access to a larger pool of computing resources than they could possibly own by themselves. Many different approaches can be used to evaluate the cloud computing service quality, and various optimization methods can be used to optimize them  \cite{6689479VMEnergy,6888495,Alicherry2013,Cohen2013,Jing2018Broker,Valerio2013SGame,He2013}.

Note that in general, there are two types of applications in the DC: service applications and batch applications \cite{6812618}. Service applications tend to generate many requests with low processing needs whereas batch applications tend to include a small number of requests with large processing needs. Unlike the batch applications that are throughput sensitive, service applications are typically response-time sensitive. In order to reduce power and energy costs, CSPs can mix online and batch tasks on the same cluster \cite{8636497}. Although the co-allocation of such tasks improves machine utilization, it challenges the DC scheduler especially for latency critical online services.

The problem under investigation in our current research is for a novel cognitive DC model wherein
\begin{enumerate}
\item a DC can determine the cost of using resources and a cloud service user can decide whether or not it will pay the price for the resource for an incoming task;
\item the online service tasks have a higher priority than batch tasks;
\item both types of tasks need a certain number of virtual machines (VM) to process.
\end{enumerate}

This paper's major contributions include:
\begin{enumerate}
\item In order to achieve the maximum total discounted expected reward for any initial state in a DC where the online services have a higher priority than the batch tasks, an Markov Decision Process model for the DCs that serving both online services and batch tasks is firstly established to gain the optimal policy in when to admit or reject a task. As far as we know, this is the first time that an cognitive DC model as described in this paper is modelled in this way with several major theoretical results obtained. This research has also significantly extend our previous work \cite{Ni8887281} with complicated challenges from non-priority treatment of the tasks in a regular DC model to the priority treatment of the task in a cognitive DC model.

\item Through a systematic probability analysis, the optimal policy in when to admit or reject a task is finally verified to be a state-related control limit policy.

\item Furthermore, to be more efficiently utilized in a cognitive DC model for the control limit policy, a lower and an upper bound, respectively, for the optimal policy values are derived theoretically also for the estimation in reality.

\item Finally, a comprehensive set of experiments on the  various cases to validate this proposed solution is conducted. As a demonstration, the machine learning method is adopted to show how to obtain the optimal values by using a feed-forward neural network model. The results offered in this paper will be expectedly utilized in various cloud data centers with different cognitive characteristics in an economically

\end{enumerate}

%\newpage
\section{Model Description and Analysis}
This section consists of two subsections, one is the description of the cognitive DC model with all needed parameters, and the other one is the establishment of the corresponding Markov decision process model along with the important components. Alternatively, the online service type is simplified as type-1 and batch task as type-2 tasks.

\subsection{The Description of the cognitive DC model}

The cognitive DC center under investigation provides two services with service type (type-1: $T_1$) and batch type (type-2: $T_2$) of application task. $T_1$ is primary user (PU) and $T_2$ is secondary user (SU), each of which will require resources in the DC. A system work flow diagram is drawn in Fig. 1.

\begin{figure}[h]
\centering
\includegraphics[width=3.5in]{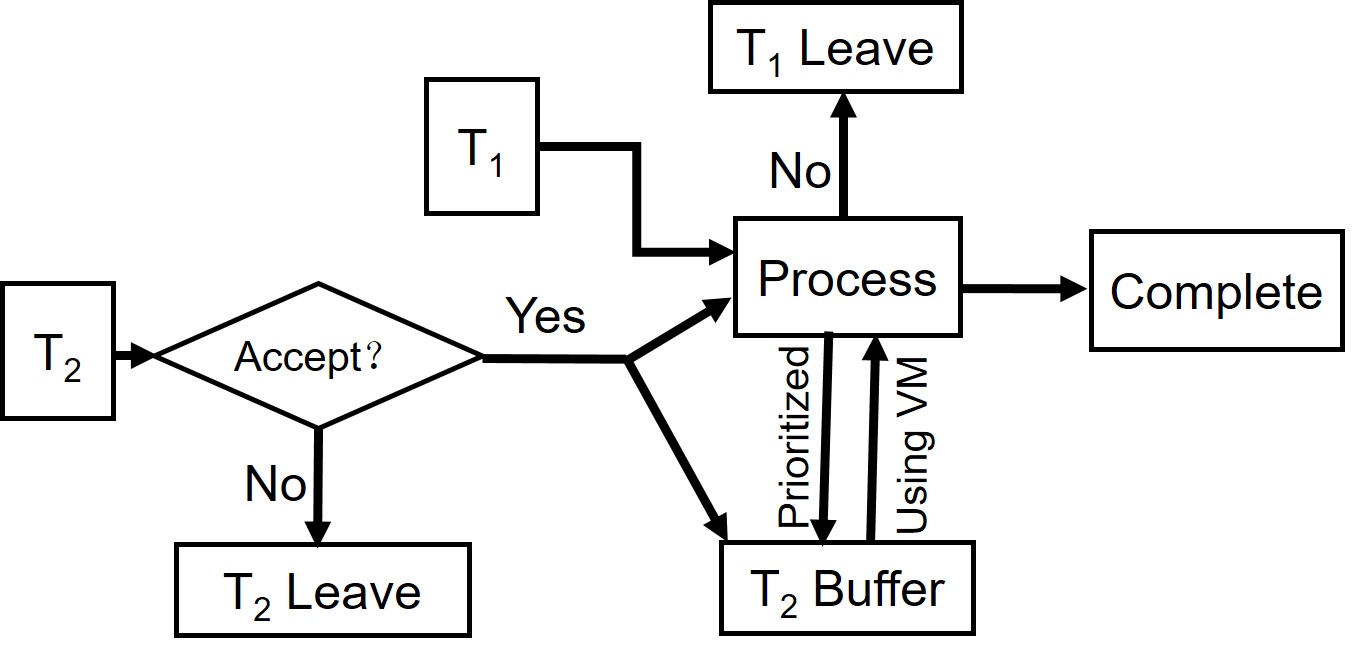}
\caption{System Work Flow.}
\label{fig_flow}
\end{figure}

The other detailed assumptions for the cognitive DC are given as follows:
\begin{enumerate}
\item There is a total number of \emph{C} VMs defining the capacity of resources from the DCs, and two types of \emph{Task}s ($T_1$ and $T_2$) in the system will share those \emph{C} VMs. $T_1$ is a type-$1$ task that is a time-sensitive service type task and needs a number of $b$ (here b is a given non-zero positive integer) VMs for service; $T_2$ is a type-$2$ task that is a specific application task and needs one VM for a sequence of operation steps. The $T_1$ task has a higher priority than the $T_2$ task as explained in the following items. Generally, since the system mainly provides service for $T_1$ tasks, it is always assumed that $C =b N_1$, where $N_1$ is a positive integer.
%\item There is a total number of \emph{C} VMs defining the capacity of resources from the DCs. Generally since the system mainly provide service for $T_1$ task, $C =b N_1$, where $N_1$ is a positive integer.

\item The arriving process for tasks $T_1$ and $T_2$ are Poisson processes with rates $\lambda_1$ and $\lambda_2$, respectively. The task processing time for these tasks in one VM follows a negative exponential distribution with rates $\mu_1$ and $\mu_2$, respectively. If a $T_1$ task is processed with multiple $b$ VMs, the service rate would be $b\mu_1$ for that task.

\item When a $T_2$ task comes to the system, the CSP will get it processed immediately when there is a VM available. However, if all VMs are busy, the CSP will decide whether to admit or reject the task based on the current number of type-1 and type-2 tasks in the system. The rejected tasks will leave the system. The admitted $T_2$ task will be put in a buffer waiting for the service whenever a VM is released for use. The waiting discipline in the buffer is ignored because the type-2 tasks in the buffer are indistinguishable. When a $T_2$ task completes the service, it will leave the system and the corresponding VM will be released for use by other tasks. When a $T_2$ task under processing is interrupted because of the arrival of a higher priority type-1 task, the interrupted $T_2$ task will be put in the buffer waiting for a new service whenever a VM is free to use.

\item When a $T_1$ task comes to the system, let $n_1$ be the number of $T_1$ tasks currently in the system; the CSP will take different actions in the following two cases:
\begin{enumerate}
\item If $n_1 < N_1$, provide the best service through allocating $b$ VMs for service, which includes the possibility of interrupting the service of several $T_2$ Tasks under processing;
\item If $n_1 = N_1$, reject the $T_1$ task.
\end{enumerate}

\item In this paper, we focus on the admission control of a $T_2$ task. Admitting and then serving a $T_2$ task would contribute $R$ units of reward to the CSP. However, if a VM serving $T_2$ task is preempted by a higher priority $T_1$ task, there is a cost, say $r (r\ge 0)$, for that interruption. To hold the tasks in the system, the CSP needs to pay a holding price at a rate $f(n_1,n_2)$ to manage the VMs (resources) in the DCs when there are  $n_1$ type-1 tasks in the process and $n_2$ type-2 tasks in the processing and in the buffer.
\end{enumerate}

It is noteworthy to point put that some notations above, including $T_1$, $T_2$, $\lambda_1$ and $\lambda_2$, and $\mu_1$ and $\mu_2$, are used as the same as those in our recent work in \cite{Ni8887281} so that the audience may easily link up with our recent works in this area.
%and [Dallass??]

To better understanding all major given parameters in this paper, we summarize them in the Table I below:

\begin{table}[!t]
\caption{A list of major given parameters}
\label{table1}
\centering
\begin{tabular}{|c|c|}
\hline
       C  &  Number of VMs in the DC  \\
%\hline
%        $n_i$  &  Number of type-$i$ tasks ($i=1,2$) \\
\hline
       $\lambda_i$  &  Arrival rate of $T_i$  ($i=1,2$)  \\
\hline
        $\mu_i$  &          Service rate of $TIi$ ($i=1,2$) \\

\hline
       $\alpha$  &  Continuous-time discount factor  \\
\hline
       $b$  &  Number of VMs needed for processing a $T_1$   \\

\hline
           $r$ & Interruption cost of a $T_2$ by a $T_1$ \\
\hline
        $R$  &          Reward of completing a $T_2$ \\
\hline
       $f(n_1,n_2)$  &  Holding cost rate at state $(n_1,n_2)$ \\
\hline

\end{tabular}
\end{table}

\subsection{Markov Decision Process Model and its Components Analysis }

Our objective in this research is to find the optimal policy such that the total expected discounted reward is maximized. In detail, if denote by $s_t$ the state at time $t$,  $a_t$ the action to take at state $s_t$, and $r(s_t, a_t)$ for the reward obtained when action $a_t$ is selected at state $s_t$, our objective is to find an optimal policy $\pi_{\alpha}$ that can bring the maximum total expected discounted reward $v_\alpha^\pi(s)$ as defined below for every initial state $s$.
\begin{equation}
\label{valphapis}
 v_\alpha^\pi(s)=E_s^\pi\bigg\{ \int_0^\infty e^{-\alpha t} r(s_t, a_t)dt\bigg\}.
 %\\
\end{equation}
Here, a policy $\pi$ specifies the decision rule to be used at every decision epoch, $\alpha$ is the discount factor. It gives the decision maker a prescription for action selection for any possible future system state or history.

Based on the model description and above objective, we can now establish a Markov decision process as follows:
\begin{enumerate}
\item Let's define the state space of the system operating process as $S = \{s:s=(n_1,n_2)\}$, where integers $n_1$ and $n_2$ satisfy $N_1\ge n_1 \ge 0$ and  $n_2\ge0$.
The event space is defined by $E = \{D_1, D_2, A_1, A_2\}$, where $D_1$ and $D_2$ means a $T_1$ and $T_2$ departure from the system after service, while $A_1$ means an arrival of a $T_1$ task, $A_2$ is an arrival of $T_2$ task. Since the states migration not only depends on the number of tasks in the system but also depends on the happening departure and arrival events, we will need to define a new state space as $\hat{S} = S \times E$.  By doing so a state could be generally written as
$\hat{s} = \langle s, e\rangle = \langle (n_1,n_2), e\rangle$, where $n_1$ and $n_2$ are the numbers for $T_1$ and $T_2$ tasks, $e$ stands for the event which will probably happen on state $(n_1,n_2)$, $e \in \{D_1, D_2, A_1, A_2\}$. Please be noticed that the specification of the event in this paper is one of major technical differences from that in paper \cite{Ni2009Control}, in which the event is assumed to happen before a state changes.

\item Denote by $a_C$ as the action to continue, the action space for states $\langle (n_1,n_2), D_1\rangle$ and $\langle (n_1,n_2), D_2\rangle$ in the state space is then given by
\begin{eqnarray*}
A_{\langle (n_1,n_2),D_1\rangle}=\{a_C\}, n_1 > 0; \\
A_{\langle (n_1,n_2),D_2\rangle}=\{a_C\}, n_2 > 0.
\end{eqnarray*}
Similarly, in states $\langle (n_1,n_2), A_1\rangle$ and $\langle (n_1,n_2), A_2\rangle$, if denote by
$a_R$ as the action to reject the request and $a_A$ as the action to
admit, the action space will be
\begin{eqnarray*}
A_{\langle (n_1,n_2), A_1\rangle}=\{a_R,a_A\}; \\%, n_1\ge 0, n_2\ge 0; \\
A_{\langle (n_1,n_2), A_2\rangle}=\{a_R,a_A\}. %, n_1\ge 0, n_2\ge 0.
\end{eqnarray*}
%In this model, we focus on the admission control for the tasks.
\item Let $C_1(n_1)$ be the number of VMs occupied by $T_1$ tasks, $C_2(n_1,n_2)$ be the number of VMs occupied by $T_2$ tasks, $C_v(n_1,n_2)$ be the number of VMs serving $T_2$ tasks that will be pre-empted if admitting a $T_1$ task, sometimes simplified as $C_1$, $C_2$ and $C_v$ in this paper. From these definitions, it is easy to know that
\begin{eqnarray*}
C_1(n_1) & =& bn_1, n_1 \le N_1,\\
C_2(n_1,n_2) & =& \min{(C-C_1(n_1), n_2)}, \\
C_v(n_1,n_2) & =& \max{(C_1+C_2+b-C, 0)} , n_1 < N_1, \\
%C_v(n_1,n_2) & =& C_2, n_1 = N_1 - 1 ;\\
C_v(n_1,n_2) & =& 0, n_1 = N_1.
\end{eqnarray*}
The decision epochs are those time points when a call arriving
or leaving the system. Based on our assumption, it is not too hard to know that the distribution of time between two epochs is
$$F(t|\hat{s},a) = 1 - e^{-\beta (\hat{s},a)t}, t\ge 0,$$
where $\hat{s} = \langle(( n_1,n_2)), b\rangle$. Denote by $s=(n_1,n_2)$ and
$\beta_0(s) = \lambda_1+\lambda_2+C_1\mu_1+C_2\mu_2.$
Since a departure event only happens when there is a task in the system, the $\beta(\hat{s},a)$ will be represented for an action $a$ as
\small
\begin{eqnarray*}
\left\{\begin{array}{ll}
            \beta_0(s) - b\mu_1, & e = D_1, a = a_C,
             n_1>0, C_2 = n_2 \\
            \beta_0(s) - b\mu_1 + & e = D_1, a = a_C, n_1>0, n_2 > C_2,\\
            \min{(n_2-C_2, b)}\mu_2, &    \\
            %& \\
            \beta_0(s) - \mu_2, & e = D_2, a = a_C, C_2=n_2>0,\\
            \beta_0(s) , & e = D_2, a = a_C, n_2>C_2,\\
            %& n_1>0, C_1+C_2 \le C\\
            \beta_0(s) + b\mu_1, & e = A_1, a = a_A,
              C_1+C_2 \le C - b,\\
            \beta_0(s) + b\mu_1 - & e = A_1, a = a_A, \\
            C_v \mu_2, & C_1+C_2 > C - b, n_1 <  N_1 - 1, \\
            \lambda_1+\lambda_2+C\mu_1, & e = A_1, a = a_A,  n_1 = N_1 - 1,\\
            \beta_0(s) + \mu_2, & e = A_2, a = a_A, C_1+C_2 < C,\\
            \beta_0(s), & e = A_2, a = a_A,  C_1+C_2 = C, \\
            %& bn_1+b_2n_2 \leq C-b_2,\\
            \beta_0(s), & e = \{A_1, A_2\}, a = a_R.  %\\
            %& n_1\ge 0, n_2\ge 0.
            \end{array}\right.
\end{eqnarray*}
\normalsize
\item Let $q(j|\hat{s},a)$ denote the probability that the system occupies
state $j$ in the next epoch, if at the current epoch the system is
at state $\hat{s}$ and the decision maker takes action $a\in A_{\hat{s}}$. For the cases
of departure events, e.g. for a departure event of $D_1$ under the condition of ($n_1>0$), $(\hat{s},a)= ( \langle(n_1,n_2), D_1\rangle, a_C)$, if denote by $s_{n_1}=(n_1-1,n_2)$, then we will have $q(j|\hat{s},a)$ as
\begin{eqnarray*}
\left\{\begin{array}{ll}
            \lambda_1/\beta_0(s_{n_1}), & j = \langle (n_1-1,n_2), A_1\rangle,\\
            \lambda_2/\beta_0(s_{n_1}), & j = \langle (n_1-1,n_2), A_2\rangle,\\
            C_1(n_1-1)\mu_1/\beta_0(s_{n_1}), & j = \langle (n_1-1,n_2), D_1\rangle,\\
            C_2(n_1-1,n_2)\mu_2/\beta_0(s_{n_1}), & j = \langle (n_1-1,n_2), D_2\rangle.\\
            \end{array}\right.
\end{eqnarray*}
Similar equations can be derived for case when $(\hat{s},a)= ( \langle(n_1,n_2), D_2\rangle, a_C)$. If denote by $s_{n_2}=(n_1,n_2-1)$ at the condition of $n_2>0$, we have
$q(j|\hat{s},a)$ as
\begin{eqnarray*}
\left\{\begin{array}{ll}
            \lambda_1/\beta_0(s_{n_2}), & j = \langle (n_1,n_2-1), A_1\rangle,\\
            \lambda_2/\beta_0(s_{n_2}), & j = \langle (n_1,n_2-1), A_2\rangle,\\
            C_1(n_1)\mu_1/\beta_0(s_{n_2}), & j = \langle (n_1,n_2-1), D_1\rangle,\\
            C_2(n_1,n_2-1)\mu_2/\beta_0(s_{n_2}), & j = \langle (n_1,n_2-1), D_2\rangle.\\
            \end{array}\right.
\end{eqnarray*}
For the cases of arrival events, such as $(\hat{s},a)= ( \langle(n_1,n_2), A_1\rangle, a_A)$, and $(\hat{s},a)= ( \langle(n_1,n_2), A_2\rangle, a_A)$, since admitting an incoming call migrates the system state immediately (adding one user or not), we will get $q(j|\hat{s},a)$ as\
\begin{eqnarray*}
 \left\{\begin{array}{ll}
            q(j|\langle (n_1+2,n_2), D_1\rangle, a_C), & e = A_1, a = a_A,\\
            q(j|\langle (n_1+1,n_2), D_1\rangle, a_C), & e = A_1, a = a_R,\\
            q(j|\langle (n_1,n_2+2), D_2\rangle, a_C), & e = A_2, a = a_A,\\
            q(j|\langle (n_1,n_2+1), D_2\rangle, a_C), & e = A_2, a = a_R.
            \end{array}\right.
\end{eqnarray*}
\item Because the system state does not change between decision epochs,
from \emph{Chp 11.5.2} ~\cite{Puterman2005MDP} and our assumptions, the expected
discounted reward between epochs satisfies
\begin{eqnarray*}
r(\hat{s},a) & = & k(\hat{s},a) + c(\hat{s},a)E_{\hat{s}}^a \left\{\int_0^{\tau_1} e^{-\alpha t}dt\right\} \nonumber \\
   & = & k(\hat{s},a) + c(\hat{s},a)E_{\hat{s}}^a \left\{[1-e^{-\alpha\tau_1}]/\alpha\right\} \nonumber \\
   & = &k(\hat{s},a) + \frac{c(\hat{s},a)}{\alpha+\beta(\hat{s},a)},
\end{eqnarray*}
where
\begin{eqnarray*}
k(\hat{s},a)=\left\{\begin{array}{cc}
 0,&e = \{D_1,D_2\}, a=a_C,\\
 0,&e = \{A_1,A_2\}, a=a_R,\\
 -C_v r,&e = A_1, a=a_A, \\
 R,&e = A_2, a=a_A.
 \end{array}\right.
\end{eqnarray*}

Also, we have the cost function $c(\hat{s},a)$ as
%\small
\begin{eqnarray*}
\left\{\begin{array}{cl}
            -f(n_1-1, n_2), & e = D_1, a = a_C, n_1 > 0, \\
            -f(n_1, n_2-1), & e = D_2, a = a_C, n_2 > 0, \\
            -f(n_1+1, n_2), & e = A_1, a = a_A,  n_1 < N_1,\\% n_1 \ge 0, n_2\ge 0, \\
            -f(n_1,n_2+1), & e = A_2, a = a_A, \\
            %& n_1 = \lceil N_1\rceil ,
            -f(n_1,n_2), & e = \{A_1, A_2\}, a=a_R.
            \end{array}\right.
\end{eqnarray*}

\end{enumerate}

In the next section we will prove that there exists a state-related threshold for accepting the tasks if the cost function has some special properties.
%\newpage
\section{Optimal Stationary State-related Control Limit Policy}

A policy is stationary if, for each decision epoch $t$, decision rule at $t$ epoch $d_t = d$ is the same.
Furthermore, a policy is called a control limit policy (or a threshold policy) for a given number of {\bf Tasks} $n_1$ and $n_2$ in the system, for $T_2$ task, is there a constant or threshold $D(n_1)\ge 0$ such that the system will accept the {\bf  arriving $T_2$} whenever the number of {\bf $T_2$} currently in the system is less than $D(n_1)$, that means the decision rule for $T_2$ is:
\begin{eqnarray}
\label{Control_Limit}
d(n_1, n_2) =\left\{\begin{array}{cc}
            Admit, & n_2 \le D(n_1),\\
            Reject, & n_2 > D(n_1).
            \end{array}\right.
\end{eqnarray}

%\subsection{Rate Uniformization and Probability Analysis}
\subsection{Total Discounted Reward}

Denote by a constant $c = \lambda_1 + \lambda_2 + C*\max(\mu_1,\mu_2)$, which is bigger than $[1-q(\hat{s}|\hat{s},a)]\beta(\hat{s},a)$. From \emph{Chp 11.5.2} ~\cite{Puterman2005MDP}, we know that there is a unique optimal solution of our model and this solution is also a stationary state-related control limit policy satisfying
\begin{equation}
\label{v(s)} v_{\alpha}^{d}(\hat{s})= r_d(\hat{s}) +
\frac{\beta_d(\hat{s})}{\alpha+\beta_d(\hat{s})} \sum_{j\in
\hat{S}}q_d(j|\hat{s})v_{\alpha}^{d}(j).
\end{equation}
By using above equation and also the uniformization technique described in ~\cite{Puterman2005MDP}, we have
\begin{eqnarray}
\label{v01}
& &  v(\langle(n_1+1,n_2),D_1\rangle)    \nonumber \\
&=&  \frac{1}{\alpha + c}
\Big[  -f(n_1, n_2)  \nonumber \\
& & \hspace{0.5cm} + \lambda_1 v(\langle(n_1,n_2),A_1\rangle) + \lambda_2 v(\langle(n_1,n_2),A_2\rangle)
 \nonumber \\
%& &+ \lambda_2 v(\langle(n_1,n_2),A_2\rangle) \nonumber \\
& & \hspace{0.5cm} + C_1(n_1)\mu_1 v(\langle (n_1,n_2), D_1\rangle) \nonumber \\
& &\hspace{0.5cm}  +  C_2(n_1,n_2)\mu_2 v(\langle (n_1,n_2), D_2\rangle) \nonumber \\
   & &\hspace{0.5cm}
    + (c-\beta_0(n_1,n_2)) v(\langle( n_1+1,n_2), D_1\rangle)
 \Big].
\end{eqnarray}
This means that
\begin{eqnarray}
\label{vv}
& &  v(\langle(n_1+1,n_2),D_1\rangle)    \nonumber \\
&=&  \frac{1}{\alpha + \beta_0(n_1, n_2) }
\Big[-f(n_1,n_2)  \nonumber \\
& &\hspace{0.5cm} + \lambda_1 v(\langle(n_1,n_2),A_1\rangle) + \lambda_2 v(\langle(n_1,n_2),A_2\rangle)
 \nonumber \\
& & \hspace{0.5cm} +  C_1(n_1)\mu_1 v(\langle (n_1,n_2), D_1\rangle) \nonumber \\
& &\hspace{0.5cm} +  C_2(n_1,n_2)\mu_2 v(\langle (n_1,n_2), D_2\rangle)
\Big].
\end{eqnarray}

Similarly, it is easily found that $$
v(\langle(n_1+1,n_2),D_1\rangle) = v(\langle(n_1,n_2+1),D_2\rangle),
$$ which shows the equality between different departure events.
%similar results can also be seen among arrival events or even between departure and arrival events.
This leads us to define a new function $X(n_1,n_2)$ as below:
\begin{eqnarray}
\label{X1}
%X(-t1,n_2)&=& v(\langle(0,n_2),D_1\rangle), \\
%X(n_1,-1)&=& v(\langle(n_1,0),D_2\rangle), \\
 X(n_1,n_2) &=& v(\langle(n_1+1,n_2),D_1\rangle) \\
             &=& v(\langle(n_1,n_2+1),D_2\rangle),
\end{eqnarray}
 for any $n_1\ge 0$ and $n_2\ge 0$.

It is noticed that $X(n_1, n_2)$,
%or sometime using $X(s)$ only for a short expression,
is only related to the state, but not with the happening event. This observation will greatly simplify the needed proof processes in next several sections.

Similar as above results for a departure event, we can consider an arrival event and will get the following results:
%Similarly, we will have
\begin{eqnarray}
\label{vA2aA}
& &  v(\langle(n_1,n_2),A_2\rangle, a_A)    \nonumber \\
&=&  R\frac{\alpha+\beta_0(n_1, n_2+1)}{\alpha + c}  \nonumber \\
&& +\frac{1}{\alpha + c} \Big[-f(n_1, n_2+1)  \nonumber \\
 & & + \lambda_1 v(\langle(n_1,n_2+1),A_1\rangle) + \lambda_2 v(\langle(n_1,n_2+1),A_2\rangle)
 \nonumber \\
& &  + C_1(n_1)\mu_1 v(\langle (n_1,n_2+1), D_1\rangle) \nonumber \\
& & + C_2(n_1, n_2+1)\mu_2 v(\langle (n_1,n_2+1), D_2\rangle) \nonumber \\
   & & + (c-\beta_0(n_1, n_2+1)) v(\langle(n_1,n_2),A_2\rangle)
\Big],
\end{eqnarray}
and
\begin{eqnarray}
\label{vA2aR}
& &  v(\langle(n_1,n_2),A_2\rangle, a_R)    \nonumber \\
&=&  \frac{1}{\alpha + c}
\Big[-f(n_1, n_2) + \lambda_1 v(\langle(n_1,n_2),A_1\rangle) \nonumber \\
& & \hspace{0.5cm}+ \lambda_2 v(\langle(n_1,n_2),A_2\rangle)
 \nonumber \\
%& &+ \lambda_2 v(\langle(n_1,n_2),A_2\rangle) \nonumber \\
& &  \hspace{0.5cm}+ C_1(n_1)\mu_1 v(\langle (n_1,n_2), D_1\rangle) \nonumber \\
&&\hspace{0.5cm} +  C_2(n_1, n_2)\mu_2 v(\langle (n_1,n_2), D_2\rangle) \nonumber \\
   & & \hspace{0.5cm}+ (c-\beta_0(n_1, n_2)) v(\langle( n_1,n_2), A_2\rangle)
\Big].
\end{eqnarray}

From above equations, we can easily get %, n_1 < \lceil N_1\rceil
\begin{eqnarray*}
v(\langle(n_1,n_2),A_2\rangle, a_A) & \geq & R + X((n_1,n_2+1)), \\
v(\langle(n_1,n_2),A_2\rangle, a_R) & \geq & X((n_1,n_2)).
\end{eqnarray*}
In fact, these two inequalities will be the equalities when the corresponding action $a_A$ or $a_R$ is the best action, respectively. %This also includes the situation when $n_1+n_2=C$, the action $a_R$ is the the best action for any arrival of $A_1$ and $A_2$.
From these analysis, it is not too hard to verify that
\begin{eqnarray}
\label{va2}
& & v(\langle(n_1,n_2),A_2\rangle)  \nonumber \\
& = &\max \Big[X((n_1,n_2)), R + X((n_1,n_2+1)) \Big].
\end{eqnarray}

For the $T_1$ tasks, as the system always accepts them until all VMs are being used, we have
\begin{eqnarray}
\label{va1}
& & v(\langle(n_1,n_2),A_1\rangle) \nonumber \\
&=&  \left\{\begin{array}{cl}
 -C_v r + X((n_1+1,n_2)), & n_1 <N_1,\\
  X((n_1,n_2)), & n_1 = N_1.
            \end{array}\right.
\end{eqnarray}

%\newpage

%Before proceeding to our major theory of optimal strategy, we will need to introduce the following two lemmas first.

\subsection{Optimal Result}

Before providing our major optimal result, we need to introduce a general result as below.
\\

{\bf Lemma 1:} Let $h(i)$ ($i \ge 0$) be an integer concave function, and denote by
$$
g(i) \equiv \max\{h(i), R + h(i+1)\} \,\,\,\, i \ge 0,
$$
for a given constant $R$. Then, $g(i)$ is also an integer concave function for $i \geq 0 $.

\emph{Proof:} Denote by $\Delta h(i) = h(i+1) - h(i)$ for any integer $i\ge 0$,  we will prove this Lemma by considering the following three cases:
%Consider $i = 1,\ldots,N,\ldots$, $N$ is a positive integer and $N > 1$ we show that $g(i)$ also is concave and nonincreasing.
%\begin{enumerate}

{\bf Case 1:} If for any $i \ge 0 $ , $\Delta h(i) \le -R$. Thus, $g(i) \equiv h(i)$ and $g(i)$ is then concave.

{\bf Case 2:} If for any $i \ge 0$, $\Delta h(i) \ge -R$. Thus, $g(i) \equiv R + h(i+1)$ and $g(i)$ is then concave.

{\bf Case 3:} We now consider the case when both of above cases would not be true. In this case, we can first know that $\Delta h(1) \ge -R$. In fact, if it is not sure, i.e., if $\Delta h(1) \le -R$, we can inductively verify that $\Delta h(i) \le -R$ for any $i\ge 0$ by noting the assumption that $h(i)$ is an integer concave function.

Since $\Delta h(1) \ge -R$, $\Delta h(i)$ is decreasing because of the concavity of $h(i)$, and the Case 2 will not hold for all $i$ in this Case 3, there then must exist an integer $k\ge 1$ such that
\begin{eqnarray*}
\Delta h(j) &\ge & -R ,\,\,\,\,\, {\rm for}\,\,\, j=1,2,...,k,    \\
\Delta h(k+1) &\le & -R.
\end{eqnarray*}
From this analysis, we will know that for any $i \geq 0$,
\begin{eqnarray*}
g(i) = \left\{\begin{array}{cl}
            R + h(i+1), & i \le k,\\
            h(i), & i > k.
            \end{array}\right.
\end{eqnarray*}
The concavity of $g(i)$, i.e., $\Delta g(i)\leq 0$, is then verified by the concavity of $h(i)$, of a constant $-R$, and the following further notification for any $i \ge 1$:
\begin{eqnarray*}
\Delta g(i) = \left\{ \begin{array}{cl}
            \Delta h(i+1), & i < k,\\
            -R, & i = k, \\
            \Delta h(i), & i > k.
            \end{array}\right.
\end{eqnarray*}

{\bf Remark 1:} It is worthy to point out that there is no condition for the given constant in the Lemma 1. This constant could be either negative or positive. In fact, we introduced this result in paper \cite{Chao1997Control} for the case when the constant is non-positive, and in paper \cite{Ni8887281} for the case when the constant is non-negative. However, it is the first time in this paper that we point out this general result without any condition on the constant $R$ and also provide the mathematical verification.

Based on above Lemma 1 and the expression developed before Lemma 1, we can now provide and then verify the following major optimal result:
%then from the equations (\ref{va1}) and (\ref{va2}), we have the following theorem.
\\

{\bf Theorem 1:}
If $f(n_1, n_2)$ is convex and increasing function on $n_2$ for any given $n_1$, the optimal policy is then a control limit policy. That means, for any state $(n_1, n_2)$, there must exist an integer, say $N_2$, such that decision
\begin{eqnarray}
\label{threshold2}
%& &
a_{\langle (n_1,n_2),A_2\rangle}
= \left\{
\begin{array}{ll}
a_A, & \mbox{if} \,\,  n_2  \leq N_2,\\
a_R, & \mbox{if} \,\,  n_2  > N_2.\\
\end{array}
\right.
\end{eqnarray}

\emph{Proof:} If all VMs are busy when an SU arrives at state $(n_1, n_2)$, we know that $C_1(n_1) + n_2 \ge C $ and then
$$C_2(n_1, n_2)=C-C_1(n_1).$$
Therefore
\begin{eqnarray}
\label{betaNon_2}
& & \beta_0(n_1, n_2+2) = \beta_0(n_1, n_2+1) = \beta_0(n_1, n_2) \nonumber \\
&=&  \lambda_1+\lambda_2+C_1(n_1)\mu_1+(C-C_1(n_1)) \mu_2,
\end{eqnarray}
which is independent of $n_2$. Further, by using the notation of $X((n_1,n_2))$, we can rewrite the equation (\ref{vv}) as below:
\begin{eqnarray}
\label{XXGC}
& &  X((n_1,n_2))     \nonumber\\
&=&  \frac{1}{\alpha + \beta_0(n_1, n_2) }
\Big[-f(n_1, n_2) + \lambda_1 v(\langle(n_1,n_2),A_1\rangle) \nonumber\\
& & + \lambda_2 v(\langle(n_1,n_2),A_2\rangle)  + C_1(n_1)\mu_1  X((n_1-1,n_2))
 \nonumber \\
& &  +(C - C_1(n_1))\mu_2   X((n_1,n_2-1))
\Big].
\end{eqnarray}
For any two-dimensional integer function $g(n_1, n_2)$ ($n_1 \ge 0, n_2 \ge 0$), we introduce the following definitions for $n_1$ and $n_2$, respectively:
\small
\begin{eqnarray}
\label{Delta1and2}
%\Delta_{n_1} g(n_1, n_2) &=& g(n_1+1, n_2) - g(n_1, n_2);\\
\Delta_{n_2} g(n_1, n_2) &=& g(n_1, n_2+1) - g(n_1, n_2).\\
\Delta^{(2)}_{n_2} g(n_1, n_2) &=& \Delta_{n_2}g(n_1, n_2+1) - \Delta_{n_2}g(n_1, n_2).
\end{eqnarray}
\normalsize

From the observation in equation (\ref{betaNon_2}) and the equation (\ref{XXGC}), we will have
\begin{eqnarray}
\label{Xcon1}
& &  \big(\alpha + \beta_0(n_1, n_2+1)\big)  \Delta_{n_2} X(n_1, n_2)  \nonumber \\
&=&
  -\Delta_{n_2}f(n_1, n_2)  \nonumber \\
   && + \lambda_1  \Delta_{n_2}  v(\langle(n_1,n_2),A_1\rangle) + \lambda_2  \Delta_{n_2}  v(\langle(n_1,n_2),A_2\rangle)  \nonumber \\
& & +  C_1(n_1)\mu_1  \Delta_{n_2} X(n_1-1,n_2)  \nonumber\\
& & + (C-C_1)\mu_2   \Delta_{n_2} X(n_1,n_2-1). %\\
\end{eqnarray}
Next, by noting
\begin{eqnarray}
\label{Cb}
&& C_v(n_1, n_2+2) = C_v(n_1, n_2+1) = C_v(n_1, n_2) \nonumber \\
&=&
\begin{array}{l}
\left\{\begin{array}{lc}
 b , & n_1 <N_1,\\
  0, &  n_1 = N_1,
            \end{array}\right.
\end{array}
\end{eqnarray}
and  equation (\ref{va1}), we will have
\begin{eqnarray}
\label{vA_1}
& & \Delta_{n_2}
v(\langle(n_1,n_2),A_1\rangle) \nonumber  \\
&=&
\begin{array}{l}
\left\{\begin{array}{cc}
   \Delta_{n_2} X(n_1+1,n_2) ,& n_1<N_1,\\
   \Delta_{n_2} X(n_1,n_2) ,& n_1=N_1.
 \end{array}\right.
\end{array}
\end{eqnarray}
By a similar implementation on about two equations (\ref{Xcon1}) and (\ref{vA_1}) by using the results in equations (\ref{betaNon_2}) and (\ref{Cb}), we have
\begin{eqnarray}
\label{Xconcave1}
& &  (\alpha + \beta_0(n_1, n_2+2))
\Delta^{(2)}_{n_2} X(n_1, n_2)   \nonumber \\
&=&
-\Delta^{(2)}_{n_2} f(n_1, n_2)  \nonumber \\
&&  + \lambda_1 \Delta^{(2)}_{n_2} v(\langle(n_1,n_2),A_1\rangle)  + \lambda_2 \Delta^{(2)}_{n_2} v(\langle(n_1,n_2),A_2\rangle) \nonumber \\
&  &+  C_1(n_1)\mu_1 \Delta^{(2)}_{n_2} X(n_1-1,n_2)  \nonumber\\
&  &+  (C-C_1)\mu_2  \Delta^{(2)}_{n_2} X(n_1,n_2-1) .%\\
\end{eqnarray}
and
\begin{eqnarray}
\label{DvA_1}
&&\Delta^{(2)}_{n_2}
v(\langle(n_1,n_2),A_1\rangle) \nonumber \\
&=&
\begin{array}{l}
\left\{\begin{array}{cc}
   \Delta^{(2)}_{n_2} X(n_1+1,n_2) ,& n_1<N_1,\\
   \Delta^{(2)}_{n_2} X(n_1,n_2) ,& n_1=N_1.
 \end{array}\right.
\end{array}
\end{eqnarray}

With the preparations on all equations from equation (\ref{XXGC}) to (\ref{DvA_1}), we can now use Value Iteration Method with three steps to show that for all states $X(n_1,n_2)$ is concave and nonincreasing for nonnegative integer function on $n_2$ for any given $n_1$ as below:

{\bf Step 1:} Set $X^{(0)}(n_1,n_2) = 0$, by noting equations (\ref{va1}) and (\ref{va2}), we know $v^{(0)}(\langle(n_1,n_2),A_2\rangle)=R$ and
\begin{eqnarray*}
%\label{va1b}
 v^{(0)}(\langle(n_1,n_2),A_1\rangle)
=  \left\{\begin{array}{cl}
 -br , & n_1 <N_1,\\
  0, &  n_1 = N_1.
            \end{array}\right.
\end{eqnarray*}
Substitute these  three results into equation (\ref{v01}), we will have
\begin{eqnarray*}
& & X^{(1)}(n_1,n_2) =  \left\{\begin{array}{cl}
 \frac{-f(n_1, n_2) +\lambda_1(-br)+\lambda_2 R }{\alpha + c} , & n_1 <N_1,\\
  \frac{-f(n_1, n_2) +0+\lambda_2 R}{\alpha + c} , &  n_1 = N_1.
            \end{array}\right.
\end{eqnarray*}
Therefore, for any $n_1$, $X^{(1)}(n_1,n_2)$ is concave and nonincreasing on $n_2$.

{\bf Step 2:}  By using above concavity and non-increasing property of $X^{(1)}(n_1,n_2)$, and the equation (\ref{va1}) for the case when all VMs are busy for state $(n_1,n_2)$, or equations (\ref{vA_1}) and (\ref{DvA_1}), we know that $v^{(1)}(\langle n_1,n_2, A_1\rangle)$ is concave and non-increasing functions for any $n_2$. By further applying the result in Lemma 1, we know that $v^{(1)}(\langle n_1,n_2, A_2\rangle)$ is also concave and non-increasing functions for any $n_2$. With these results in mind, and using the results in equations (\ref{Xcon1}) and (\ref{Xconcave1}), we will know that
$$
\Delta_{n_2} X^{(2)}(n_1, n_2) \leq 0, \hspace{0.5cm} {\rm and} \hspace{0.5cm}
\Delta^{(2)}_{n_2} X^{(2)}(n_1, n_2) \leq 0. $$
   These two inequalities justify that for any $n_1$,
    $X^{(2)}(n_1,n_2)$ is  nonincreasing and concave on $n_2$.

%\item
{\bf Step 3:} Finally, by noting the \emph{Theorem 11.3.2} of ~\cite{Puterman2005MDP} that the optimality equation has the unique solution, we know the value iteration $X^{(n)}(n_1,n_2)$ will uniquely converges. Therefore, as the iteration continues, with $n$ goes to $\infty$, we know that for any $n_1$, $X(n_1,n_2)$ is always concave
nonincreasing for any $n_2$.
%\end{enumerate}

Finally, by using the equation (\ref{va2}) and the property of nonincreasing and concave on $n_2$ for $X(n_1, n_2)$, it is straight forward to know that the optimal policy is a control limit policy as stated in the Theory.

The proof is now completed.

%\newpage
\section{Bound Analysis for the Optimal Result}
From the above analysis, we have known that discounted optimal policy is a control limit policy if the cost function  $f(n_1, n_2)$ is convex and increasing on $n_2$ for any given  $n_1$. However, since identification of the optimal value is pretty important in reality, how to determine the corresponding threshold value of the control limit policy or the range of the value becomes a challenging issue as long as the optimal strategy if verified. In the Section, we will step toward to the identification of the optimal value's range and derive a useful range result in terms of the lower bound and the upper bound of the optimal value. We now first introduce a general result as below.

{\bf Lemma 2:} Let $h_k(i)$ be an integer concave function (k=1,2), and for a constant $R$ denote by
$$
g_k(i) \equiv \max\{h_k(i), R + h_k(i+1)\}, \hspace{0.5cm} k=1,2.
$$
Then, $\Delta g_1(i) \le \Delta g_2(i)$ holds if $\Delta h_1(i) \le \Delta h_2(i)$ for any $i \geq 0.$

\emph{Proof:} For any given integer $i$, since $\Delta h_1(i) \le \Delta h_2(i)$, we only need to verify the result is true for the following three cases.

{\bf Case 1:} If $\Delta h_2(i) \le -R$ is true. In this case, since $\Delta g_k(i) = \Delta h_k(i)$, it is straightforward to know
$$\Delta g_1(i) = \Delta h_1(i) \le \Delta h_2(i) = \Delta g_2(i) .$$

{\bf Case 2:} If $-R \le \Delta h_1(i)$ is true. In this case, we know $g_k(i)=R+h_k(i+1)$ (k=1, 2), and then,
\begin{eqnarray*}
\Delta g_k(i) &=&   g_k(i+1)- [R+h_k(i+1)]  \\
              &=&     \max\{-R, \Delta h_k(i+1)\}.
\end{eqnarray*}
 From above result, it is also directly to know that
 $$\Delta g_1(i) \le  \Delta g_2(i),$$
 by noting $\Delta h_1(i+1) \le \Delta h_2(i+1)$.

{\bf Case 3:} If $\Delta h_1(i) \le -R \le \Delta h_2(i)$ is true. In this case, from analysis in above Case 1 and Case 2, we know that
$\Delta g_1(i) = \Delta h_1(i)$, and
$$
\Delta g_2(i) = \max \{-R, \Delta h_2(i+1) \}.
$$
Therefore, $\Delta g_1(i) \le -R \le  \Delta g_2(i)$.

The proof is now completed.

%{\color{red} The following 4 equations are listed for easy understanding.}
With the condition that the cost function  $f(n_1, n_2)$ is convex and increasing on $n_2$ for any given $n_1$, from Theorem 1, we already know that $X(n_1, n_2)$ is concave and decreasing on $n_2$ for any $n_1$, and therefore we will have
\begin{eqnarray}
\label{Xrs1}
%\Delta_{n_2} X(n_1-1, n_2) & \ge  & \Delta_{n_2} X(n_1, n_2),\nonumber\\
\Delta_{n_2} X(n_1, n_2-1) & \ge  & \Delta_{n_2} X(n_1, n_2).
\end{eqnarray}
Also from Lemma 1 and equation (\ref{va1}) and (\ref{va2}), we can also have
\begin{eqnarray}
\label{Xrs2}
\Delta_{n_2} v(\langle(n_1,n_2),A_1\rangle) & \le  & \Delta_{n_2} X(n_1, n_2),\nonumber\\
\Delta_{n_2} v(\langle(n_1,n_2),A_2\rangle) & \le  & \Delta_{n_2} X(n_1, n_2). %,\nonumber\\
%\Delta_{n_2} X(n_1-1, n_2) & \ge  & \Delta_{n_2} X(n_1, n_2),\nonumber\\
%\Delta_{n_2} X(n_1, n_2-1) & \ge  & \Delta_{n_2} X(n_1, n_2).
\end{eqnarray}

From these results, by noting equation (\ref{betaNon_2}), we may rewrite equation (\ref{Xcon1}) as below:

\begin{eqnarray}
\label{Xfcon1}
& &  \alpha\Delta_{n_2} X(n_1, n_2) + \Delta_{n_2} f (n_1, n_2)   \nonumber \\
&=& \hspace{0.25cm}
\lambda_1 \big[\Delta_{n_2} v(\langle(n_1,n_2),A_1\rangle) - \Delta_{n_2} X(n_1, n_2)\big] \nonumber\\
& & + \lambda_2 \big[\Delta_{n_2} v(\langle(n_1,n_2),A_2\rangle) - \Delta_{n_2}X(n_1, n_2)\big] \nonumber \\
& &  + C_1(n_1)\mu_1 \big[ \Delta_{n_2} X(n_1-1,n_2) - \Delta_{n_2} X(n_1, n_2) \big] \nonumber\\
& &  + C_2\mu_2 \big[ \Delta_{n_2} X(n_1,n_2-1) - \Delta_{n_2} X(n_1,n_2)\big].%\\
\end{eqnarray}
\normalsize

We also need to introduce a result below before presenting our major boundary result:

{\bf Lemma 3:} If the cost function $f(n_1,n_2)$ is a convex and nondecreasing function on $n_2$ for any given $n_1$, and $\Delta_{n_2} f(n_1, n_2)$ is a nondecreasing function on $n_1$, then $\Delta_{n_2} X(n_1,n_2)$ a nonincreasing function on $n_1$.

\emph{Proof:} The detailed verification is included in Appendix.\\

From these preparation, we can know provide and prove our bound result as below.

{\bf Theorem 2:}
If the cost function $f(n_1,n_2)$ is a convex and nondecreasing function on $n_2$ for any given $n_1$, and $\Delta_{n_2} f(n_1, n_2)$ is a nondecreasing function on $n_1$, then we will have
\begin{enumerate}
\item A {\bf lower bound} of the optimal threshold value $N_2$ is given by
$$
N_{2*}=\max \left\{n_2: \Delta_{n_2} f(N_1, n_2)  < \alpha R     \right\},
$$
if there exists an integer $n_2$ such that
\begin{equation}
\label{ubcf1}
\Delta_{n_2} f(N_1, n_2)  < \alpha R.
\end{equation}
\item An {\bf upper bound} of the optimal threshold value $N_2$ is given by
%\begin{enumerate}
%\item when there is at least one free VM when an SU arrives
\begin{eqnarray*}
N^*_{2} =\min \Big\{  n_2: \Delta_{n_2} f(0, n_2)  \hspace{3.6cm}   \\
 \hspace{2cm}  > (\alpha+ \min(n_2+1,C)\mu_2) R  \Big\},
\end{eqnarray*}
if there exists an integer $n$ such that
\begin{equation}
\label{ubcf2.1}
\Delta_{n_2} f(0, n_2)  > (\alpha+\min(n_2+1,C)\mu_2) R.
\end{equation}
%\item when all the VMs are busy when an SU arrives
%$$
%N^*_{2}=\min \left\{n: \Delta_{n_2} f(0, n)  > (\alpha+C\mu_2) R     \right\},
%$$
%if there exists an integer $n$ such that
%\begin{equation}
%\label{ubcf2.2}
%\Delta_{n_2} f(0, n)  > (\alpha+C\mu_2) R,
%\end{equation}

%\end{enumerate}
%{\color{red}
%(This condition can be removed if we can verify that it is impossible for all n,  $\Delta_{n_2} f(0, n)  < (\alpha+C\mu_2) R.$)}
%\item The system will not accept any $T_2$ arrivals if
%$$
%\Delta_{n_2} f(0, 0)  > (\alpha+\mu_2) R
%$$
\end{enumerate}

\emph{Proof:} We will have the following justifications:
\begin{enumerate}
\item It is intuitive that an integer $n_2$ would be a lower bound for the optimal schedule value in the Theorem 1 as long as the action on the state $(n_1, n_2)$ is acceptance for an arrived SU. Next, since $N_1$ is the largest value of all $n_1$ in state $(n_1,n_2)$, an acceptance action for arrived SU at state $(N_1, n_2)$ must result in an acceptance action for arrived SU at state $(n_1, n_2)$. Therefore, the optimal threshold value in Theorem 1 for state $(N_1, n_2)$ is bigger than the optimal threshold value in Theorem 1 for state $(n_1, n_2)$. This also implies that the lower bound of optimal threshold value in Theorem 1 for any state $(N_1, n_2)$ is more closer to the optimal threshold value than that for any state $(n_1, n_2)$.

    To verify the lower bound result in the Theorem 2, we only need to verify that the action on the state $(N_1, n_2)$ is acceptance for an arrived SU under the condition in equation (\ref{ubcf1}).
     Or equivalently, we only need to verify that if the action on the state $(N_1, n_2)$ is rejection, then the equation (\ref{ubcf1}) must not be valid. In fact, at state $(N_1, n_2)$, there is no free VM for both PU and SU arrivals, which means the system will take the action to reject both PU and SU arrival. By noticing the equation (\ref{va1}) and (\ref{va2}),
 %non-increasing property of the concave functions $v(\langle n_1,n_2, A_1\rangle)$, $v(\langle n_1,n_2, %A_2\rangle)$ and $X(n_1,n_2)$,
 %suppose if equation (\ref{ubcf1}) holds,
 if system will reject the arrived PU and SU at sate $(N_1, n_2)$, we have
 \begin{eqnarray*}
\Delta_{n_2} v(\langle(N_1,n_2),A_1\rangle) & =  & \Delta_{n_2} X(N_1, n_2),\\
\Delta_{n_2} v(\langle(N_1,n_2),A_2\rangle) & =  & \Delta_{n_2} X(N_1, n_2).
\end{eqnarray*}
Submitting these results into the equation (\ref{Xfcon1}), by recalling the concavity of $X(N_1, n_2))$ on $n_2$ in the verification of Theorem 1 and the result in Lemma 3, we know the right-hand side of the equation is non-negative. Therefore, we will have
$$
\Delta_{n_2} f(N_1, n_2) \geq -\alpha\Delta_{n_2} X(N_1, n_2) \geq \alpha R.
$$
The last inequality comes from the rejection of SU at state $(N_1, n_2)$ by equation (\ref{va2}). Thus, we verified that the equation (\ref{ubcf1}) is invalid. Therefore, the verification of the low bound for any state $(n_1, n_2))$ is now completed.

\item On the contrary, an integer $n_2$ would be an upper bound for the optimal schedule value in the Theorem 1 as long as the action on the state $(n_1, n_2)$ is rejection for an arrived SU. We first consider the case of $n_2 < C$, which means there is at least one free VM when an SU arrives. Therefore, to verify the upper bound result in the Theorem 2, we only need to verify that the action on the state $(0, n_2)$ is rejection for an arrived SU. Or equivalently, we only need to verify that if the action on the state $(0, n_2)$ is acceptance, then the equation (\ref{ubcf2.1}) must not be valid. In fact, at state $(0, n_2)$, by noticing the equation (\ref{va1}) and (\ref{va2}), if system will accept the arrived SU ($T_2$) at sate $(0, n_2)$, $n_2 < C$, we have
 \begin{eqnarray*}
\Delta_{n_2} v(\langle(0,n_2),A_1\rangle) & \le  & \Delta_{n_2} X(0, n_2),\\
\Delta_{n_2} v(\langle(0,n_2),A_2\rangle) & \le  & \Delta_{n_2} X(0, n_2).
\end{eqnarray*}
Since the action is to accept $T_2$ arrival, from (\ref{va2}) we have
\begin{eqnarray*}
\Delta_{n_2} X(0, n_2-1) \ge &   \Delta_{n_2} X(0, n_2)\ge & -R.
\end{eqnarray*}
Submitting these results into the equation (\ref{Xfcon1}), by again recalling the concavity of $X(n_1, n_2))$ in the verification of Theorem 1 and the result in Lemma 3, we will have
$$
(\alpha+(n_2+1)\mu_2) \Delta_{n_2} X(0, n_2) + \Delta_{n_2} f (0, n_2) \le 0.
$$
Therefore, we will have
 \begin{eqnarray*}
\Delta_{n_2} f(0, n_2) & \leq & -(\alpha+(n_2+1)\mu_2) \Delta_{n_2} X(0, n_2) \\
 & \leq & (\alpha+(n_2+1)\mu_2) R.
\end{eqnarray*}
The last inequality comes from the acceptance of SU at state $(0, n_2)$ for $n_2 < C$ by equation (\ref{va2}). Thus, we verified that the equation (\ref{ubcf2.1}) is invalid.

For the states $(0, n_2)$ with $n \ge C$, which means all the VMs are busy when an SU arrives, a similar upper bound deduction can be derived from equation (\ref{Xfcon1}) as
\begin{equation*}
\Delta_{n_2} f(0, n_2)  > (\alpha+C\mu_2) R.
\end{equation*}
Since $f(0, n_2)$ is the smallest value of all $n_1$, the upper bound for state $(0,n_2)$ is therefore also an upper bound for any other state $(n_1,n_2)$  with $n_1 > 0$. The verification of the upper bound in the Theorem is now completed by the justification in the beginning of this paragraph.

%\item
\end{enumerate}
 {\bf Remark 2:} More specifically, if equation (\ref{ubcf1}) never holds, i.e., $\Delta_{n_2} f (n_1, n_2) > \alpha R$ holds for all states, it can be observed,  from equation (\ref{ubcf2.1}) on the case when $n_2=0$, that the upper bound becomes $0$. This means the system will not accept any $T_2$ arrivals into the system if equation (\ref{ubcf1}) never holds.

%{\bf Theorem 3:} If at least one VM is available when an SU arrives, then
%\begin{enumerate}
%\item a {\bf lower bound} of the threshold value is given by
%$$
%N_{2*}=\max \left\{n: \Delta_{n_2} f(N_1, n)  < (\alpha+\mu_2) R     \right\},
%$$
%if there exists an integer $n$ such that
%\begin{equation}
%\label{ubcf3}
%\Delta_{n_2} f(N_1, n)  < \alpha R.
%\end{equation}
%
%\item an {\bf upper bound} of the threshold value is given by
%$$
%N^*_{2}=\min \left\{n: \Delta_{n_2} f(0, n)  > (\alpha+(n+1)\mu_2) R     \right\},
%$$
%if there exists an integer $n$ such that
%\begin{equation}
%\label{ubcf4.2}
%\Delta_{n_2} f(0, n)  \ge (\alpha+(n+1)\mu_2) R.
%\end{equation}
%
%\end{enumerate}
%
%\emph{Proof:} It is seen that when $n_1=N_1$, there is no free VM for both PU and SU arrivals, but based on the observation that due to the convex and nonincreasing property of $f(n_1,n_2)$, the values of threshold is decreasing as $n_1 \uparrow$, so the lower bound from $n_1=N_1$ is also the lower bound for $n_1 < N_1$, including the cases there is free VM when an SU arrives. For the upper bound, it is smaller compared to Theorem 2, please see the verification in the Appendix C.

%\newpage
\section{Numerical Analysis}
We have theoretically verified that the optimal policy to maximize the total expected discounted reward in equation (\ref{valphapis}) is a control limit policy or a threshold policy for accepting task-2 arrivals. Our CTMDP model consists of several parameters like arrival rates, departure rates, rewards and cost function, etc. Here for simulation and numerical analysis purpose we set some parameters as those listed in TABLE II. It can be seen from the Table II that the loads for $T_1$ and $T_2$ tasks are $\rho_1=\frac{\lambda_1}{\mu_1}=1/6$ and $\rho_2=\frac{\lambda_2}{\mu_2}=1/4$, which means the system are lightly loaded. Other parameters of the system is set to be as $R = 5$, $r = 0.5$, $C=10$, $b=5$, discount factor be $\alpha = 0.1$, holding cost function be $f(x,y) = x^2 + y^2$.
%\begin{table}[!t]
%% increase table row spacing, adjust to taste
%\renewcommand{\arraystretch}{1.3}
% if using array.sty, it might be a good idea to tweak the value of
% \extrarowheight as needed to properly center the text within the cells
%\caption{An Example of a Table}
%\label{table_example}
%\centering
%% Some packages, such as MDW tools, offer better commands for making tables
%% than the plain LaTeX2e tabular which is used here.
%\begin{tabular}{|c||c|}
%\hline
%One & Two\\
%\hline
%Three & Four\\
%\hline
%\end{tabular}
%\end{table}

\begin{table}[!t]
\caption{Parameter value setting}
\label{table2}
\centering
%\begin{center}
\begin{tabular}{|c|c|c|c|}
\hline
         & $\lambda$  & $\mu$   \\
\hline
        $T_1$  &          1 &       6   \\
\hline
        $T_2$  &          2 &       8   \\
\hline
\end{tabular}
\end{table}
%\textbf{Table 2:} Parameter value setting
%\end{center}

However, it is always a challenging problem in exactly finding this threshold value or optimal objective value for a given problem, especially in a continuously changing environment, such as the changes of arrival rates, service rates, rewards in the real world. While we have demonstrated in above subsection on how to derive the thresholds by using value iteration method, the calculation through this way is always a time-consuming issue.

In the 3rd subsection, we propose the machine learning method and then demonstrate it to obtain or estimate the threshold value and the optimal objective value by using a feed-forward neural network model. A feed-forward neural network is an artificial neural network where connections between the units do not form a cycle.

\subsection{Threshold Policy}
With this parameter setting, by using value iteration method we have
\begin{table}[!t]
\caption{$X(n_1, n_2)$ Values with Optimal Policy}
\label{table3}
\centering
\begin {tabular} {|c|c|c|c|c|c|c|} 
 \hline 
  & 	0 & 	  & 	  & 	$n_2\rightarrow$ & 	  & 	5 	\\ 
 \hline 
0 & 	   96.53 & 	   96.38 & 	   96.10 & 	   95.69 & 	   95.16 & 	   94.51 	\\ 
 \hline 
$n_1\downarrow$ & 	   96.50 & 	   96.33 & 	   96.03 & 	   95.61 & 	   95.07 & 	   94.40 	\\ 
 \hline 
2 & 	   96.43 & 	   96.24 & 	   95.89 & 	   95.38 & 	   94.72 & 	   93.89 	\\ 
 \hline 
  & 	6 & 	  & 	  & 	$n_2\rightarrow$ & 	  & 	11 	\\ 
 \hline 
0 & 	   93.72 & 	   92.80 & 	   91.74 & 	   90.55 & 	   89.22 & 	   87.63 	\\ 
 \hline 
$n_1\downarrow$ & 	   93.51 & 	   92.41 & 	   91.11 & 	   89.61 & 	   87.90 & 	   85.93 	\\ 
 \hline 
2 & 	   92.81 & 	   91.49 & 	   89.94 & 	   88.14 & 	   86.11 & 	   83.78 	\\ 
 \hline 
  & 	12 & 	  & 	  & 	$n_2\rightarrow$ & 	  & 	17 	\\ 
 \hline 
0 & 	   85.73 & 	   83.51 & 	   80.95 & 	   78.00 & 	   74.66 & 	   70.89 	\\ 
 \hline 
$n_1\downarrow$ & 	   83.65 & 	   81.03 & 	   78.03 & 	   74.62 & 	   70.79 & 	   66.50 	\\ 
 \hline 
2 & 	   81.11 & 	   78.06 & 	   74.61 & 	   70.71 & 	   66.35 & 	   61.51 	\\ 
 \hline 
  & 	18 & 	  & 	  & 	$n_2\rightarrow$ & 	  & 	23 	\\ 
 \hline 
0 & 	   66.68 & 	   61.99 & 	   56.81 & 	   51.11 & 	   44.88 & 	   38.09 	\\ 
 \hline 
$n_1\downarrow$ & 	   61.73 & 	   56.46 & 	   50.68 & 	   44.34 & 	   37.44 & 	   29.94 	\\ 
 \hline 
2 & 	   56.17 & 	   50.29 & 	   43.87 & 	   36.87 & 	   29.26 & 	   21.02 	\\ 
 \hline 
\end{tabular}

%\textbf{Table 3:} $X(n_1, n_2)$ Values with Optimal Policy
\end{table}
As observed from TABLE III, $X(n_1, n_2)$ values are concave decreasing on $n_2$ direction, which fits our theoretical result.
\begin{table}[!t]
\caption{Actions for $T_2$ task of Optimal Policy}
\label{table4}
\centering
\begin {tabular} {|c|c|c|c|c|c|c|c|c|} 
 \hline 
  & 	0 & 	  & 	  & 	  & 	$n_2\rightarrow$ & 	  & 	  & 	7 	\\ 
 \hline 
0 & 	1 & 	1 & 	1 & 	1 & 	1 & 	1 & 	1 & 	1 	\\ 
 \hline 
$n_1\downarrow$ & 	1 & 	1 & 	1 & 	1 & 	1 & 	1 & 	1 & 	1 	\\ 
 \hline 
2 & 	1 & 	1 & 	1 & 	1 & 	1 & 	1 & 	1 & 	1 	\\ 
 \hline 
  & 	8 & 	  & 	  & 	  & 	$n_2\rightarrow$ & 	  & 	  & 	15 	\\ 
 \hline 
0 & 	1 & 	1 & 	1 & 	1 & 	1 & 	1 & 	1 & 	1 	\\ 
 \hline 
$n_1\downarrow$ & 	1 & 	1 & 	1 & 	1 & 	1 & 	1 & 	1 & 	1 	\\ 
 \hline 
2 & 	1 & 	1 & 	1 & 	1 & 	1 & 	1 & 	1 & 	1 	\\ 
 \hline 
  & 	16 & 	  & 	  & 	  & 	$n_2\rightarrow$ & 	  & 	  & 	23 	\\ 
 \hline 
0 & 	1 & 	1 & 	1 & 	0 & 	0 & 	0 & 	0 & 	0 	\\ 
 \hline 
$n_1\downarrow$ & 	1 & 	1 & 	0 & 	0 & 	0 & 	0 & 	0 & 	0 	\\ 
 \hline 
2 & 	1 & 	0 & 	0 & 	0 & 	0 & 	0 & 	0 & 	0 	\\ 
 \hline 
\end{tabular}

%\textbf{Table 4:} Actions for $T_2$ task of Optimal Policy
\end{table}
As noticed from TABLE IV, "1" means the system will accept the arrival, "0" means the system will reject the arrival. Since the reward $R$ is large, from the table we see the system will accept $T_2$ tasks into the buffer even there is already some $T_2$ tasks waiting.

Next we decrease the reward $R=1$, then we have the $X(n_1, n_2)$ Value as listed in TABLE V. As seen from TABLE VI, with a smaller reward, from the table we see the system will not accept many $T_2$ tasks into the waiting buffer.
\begin{table}[!t]
\caption{$X(n_1, n_2)$ Values with Optimal Policy}
\label{table5}
\centering
\begin {tabular} {|c|c|c|c|c|c|c|} 
 \hline 
  & 	0 & 	  & 	  & 	$n_2\rightarrow$ & 	  & 	5 	\\ 
 \hline 
0 & 	   16.53 & 	   16.38 & 	   16.10 & 	   15.69 & 	   15.16 & 	   14.51 	\\ 
 \hline 
$n_1\downarrow$ & 	   16.50 & 	   16.33 & 	   16.03 & 	   15.61 & 	   15.07 & 	   14.40 	\\ 
 \hline 
2 & 	   16.43 & 	   16.24 & 	   15.89 & 	   15.38 & 	   14.72 & 	   13.89 	\\ 
 \hline 
  & 	6 & 	  & 	  & 	$n_2\rightarrow$ & 	  & 	11 	\\ 
 \hline 
0 & 	   13.72 & 	   12.80 & 	   11.75 & 	   10.57 & 	    9.26 & 	    7.68 	\\ 
 \hline 
$n_1\downarrow$ & 	   13.51 & 	   12.43 & 	   11.14 & 	    9.65 & 	    7.97 & 	    6.03 	\\ 
 \hline 
2 & 	   12.83 & 	   11.52 & 	    9.99 & 	    8.22 & 	    6.22 & 	    3.94 	\\ 
 \hline 
  & 	12 & 	  & 	  & 	$n_2\rightarrow$ & 	  & 	17 	\\ 
 \hline 
0 & 	    5.82 & 	    3.64 & 	    1.12 & 	   -1.76 & 	   -5.03 & 	   -8.72 	\\ 
 \hline 
$n_1\downarrow$ & 	    3.79 & 	    1.22 & 	   -1.72 & 	   -5.05 & 	   -8.80 & 	  -12.99 	\\ 
 \hline 
2 & 	    1.32 & 	   -1.66 & 	   -5.04 & 	   -8.85 & 	  -13.11 & 	  -17.85 	\\ 
 \hline 
\end{tabular}

%\textbf{Table 5:} $X(n_1, n_2)$ Values with Optimal Policy
\end{table}

\begin{table}[!t]
\caption{Actions for $T_2$ task of Optimal Policy}
\label{table6}
\centering
\begin {tabular} {|c|c|c|c|c|c|c|c|c|} 
 \hline 
  & 	0 & 	  & 	  & 	  & 	$n_2\rightarrow$ & 	  & 	  & 	7 	\\ 
 \hline 
0 & 	1 & 	1 & 	1 & 	1 & 	1 & 	1 & 	1 & 	1 	\\ 
 \hline 
$n_1\downarrow$ & 	1 & 	1 & 	1 & 	1 & 	1 & 	1 & 	1 & 	1 	\\ 
 \hline 
2 & 	1 & 	1 & 	1 & 	1 & 	1 & 	1 & 	1 & 	1 	\\ 
 \hline 
  & 	8 & 	  & 	  & 	  & 	$n_2\rightarrow$ & 	  & 	  & 	15 	\\ 
 \hline 
1 & 	1 & 	1 & 	1 & 	0 & 	0 & 	0 & 	0 & 	0 	\\ 
 \hline 
$n_1\downarrow$ & 	1 & 	0 & 	0 & 	0 & 	0 & 	0 & 	0 & 	0 	\\ 
 \hline 
2 & 	0 & 	0 & 	0 & 	0 & 	0 & 	0 & 	0 & 	0 	\\ 
 \hline 
\end{tabular}

%\textbf{Table 6:} Actions for $T_2$ task of Optimal Policy
\end{table}

\subsection{Threshold to Values}
From equation (\ref{valphapis}), the optimal policy would get the maximum total expected discounted reward if starting from an initial state. Therefore, by using the obtained thresholds in a policy, we can also calculate the corresponding total expected discounted reward using equation (\ref{valphapis}). Using rate uniformization technique, the expected time between two epoch is $\frac{1}{c}$. The calculation runs until the discount is less than $1E-6$ and the Expected $X(n_1, n_2)$ Values are shown in TABLE VII and VIII.
\begin{table}[!t]
\caption{Expected $X(n_1, n_2)$ Values of Optimal Policy with $R=1$}
\label{table7}
\centering
\begin {tabular} {|c|c|c|c|c|c|c|} 
 \hline 
  & 	0 & 	  & 	  & 	$n_2\rightarrow$ & 	  & 	5 	\\ 
 \hline 
0 & 	   16.53 & 	   16.38 & 	   16.10 & 	   15.69 & 	   15.16 & 	   14.51 	\\ 
 \hline 
$n_1\downarrow$ & 	   16.50 & 	   16.33 & 	   16.03 & 	   15.61 & 	   15.07 & 	   14.40 	\\ 
 \hline 
2 & 	   16.43 & 	   16.24 & 	   15.89 & 	   15.38 & 	   14.72 & 	   13.89 	\\ 
 \hline 
  & 	6 & 	  & 	  & 	$n_2\rightarrow$ & 	  & 	11 	\\ 
 \hline 
0 & 	   13.72 & 	   12.80 & 	   11.75 & 	   10.57 & 	    9.26 & 	    7.68 	\\ 
 \hline 
$n_1\downarrow$ & 	   13.51 & 	   12.43 & 	   11.14 & 	    9.65 & 	    7.97 & 	    6.03 	\\ 
 \hline 
2 & 	   12.83 & 	   11.52 & 	    9.99 & 	    8.22 & 	    6.22 & 	    3.94 	\\ 
 \hline 
  & 	12 & 	  & 	  & 	$n_2\rightarrow$ & 	  & 	17 	\\ 
 \hline 
0 & 	    5.82 & 	    3.64 & 	    1.12 & 	   -1.76 & 	   -5.03 & 	   -8.72 	\\ 
 \hline 
$n_1\downarrow$ & 	    3.79 & 	    1.22 & 	   -1.72 & 	   -5.05 & 	   -8.80 & 	  -12.99 	\\ 
 \hline 
2 & 	    1.32 & 	   -1.66 & 	   -5.04 & 	   -8.85 & 	  -13.11 & 	  -17.85 	\\ 
 \hline 
\end{tabular}

%\textbf{Table 7:} Expected $X(n_1, n_2)$ Values of Optimal Policy with $R=1$
\end{table}

\begin{table}[!t]
\caption{Expected $X(n_1, n_2)$ Values of Optimal Policy with $R=5$}
\label{table8}
\centering
\begin {tabular} {|c|c|c|c|c|c|c|} 
 \hline 
  & 	0 & 	  & 	  & 	$n_2\rightarrow$ & 	  & 	5 	\\ 
 \hline 
0 & 	   96.53 & 	   96.38 & 	   96.10 & 	   95.69 & 	   95.16 & 	   94.51 	\\ 
 \hline 
$n_1\downarrow$ & 	   96.50 & 	   96.33 & 	   96.03 & 	   95.61 & 	   95.07 & 	   94.40 	\\ 
 \hline 
2 & 	   96.43 & 	   96.24 & 	   95.89 & 	   95.38 & 	   94.72 & 	   93.89 	\\ 
 \hline 
  & 	6 & 	  & 	  & 	$n_2\rightarrow$ & 	  & 	11 	\\ 
 \hline 
0 & 	   93.72 & 	   92.79 & 	   91.74 & 	   90.55 & 	   89.22 & 	   87.63 	\\ 
 \hline 
$n_1\downarrow$ & 	   93.51 & 	   92.41 & 	   91.11 & 	   89.61 & 	   87.90 & 	   85.93 	\\ 
 \hline 
2 & 	   92.81 & 	   91.49 & 	   89.94 & 	   88.14 & 	   86.11 & 	   83.78 	\\ 
 \hline 
  & 	12 & 	  & 	  & 	$n_2\rightarrow$ & 	  & 	17 	\\ 
 \hline 
0 & 	   85.73 & 	   83.51 & 	   80.95 & 	   78.00 & 	   74.66 & 	   70.89 	\\ 
 \hline 
$n_1\downarrow$ & 	   83.65 & 	   81.03 & 	   78.03 & 	   74.62 & 	   70.79 & 	   66.49 	\\ 
 \hline 
2 & 	   81.11 & 	   78.06 & 	   74.61 & 	   70.71 & 	   66.35 & 	   61.51 	\\ 
 \hline 
  & 	18 & 	  & 	  & 	$n_2\rightarrow$ & 	  & 	23 	\\ 
 \hline 
0 & 	   66.67 & 	   61.98 & 	   56.81 & 	   51.11 & 	   44.88 & 	   38.09 	\\ 
 \hline 
$n_1\downarrow$ & 	   61.73 & 	   56.46 & 	   50.67 & 	   44.34 & 	   37.44 & 	   29.94 	\\ 
 \hline 
2 & 	   56.17 & 	   50.29 & 	   43.87 & 	   36.86 & 	   29.26 & 	   21.02 	\\ 
 \hline 
\end{tabular}

%\textbf{Table 8:} Expected $X(n_1, n_2)$ Values of Optimal Policy with $R=5$
\end{table}
Comparing Table VII and VIII to the TABLE III and V, we see the differences between these Tables is very small, which proves the correctness of the calculation from the Value Iteration method in the last subsection.

\subsection{Threshold Estimation with Machine Learning}
Taking the parameters in the CTMDP model as the input and the corresponding threshold value as the output, we can build the neural network model as shown in Fig. 2. Based on these development, we can build the training data sets as shown in TABLE IX, here reward $R$ is chosen from $1,\,2,\,3,\,4,\,5,\,6,\,7,\,8$, $\lambda_2$ is chosen from $1, \,2, \,3, \,4, \,5$, and $\mu_2$ is chosen from $8, \,10, \,12, \,14, \,16$, while keeping all the other parameters unchanged.

The combination of these parameters will give us $8*5*5=200$ training data inputs totally. More specifically, as shown in Fig. 2, this is a two-layer neural network with 30 hidden neurons. The inputs are the set of parameters in our CTMDP model, such as reward, arrival rates, departure rates, which makes the total input parameters be 5; output are the threshold values for each $n_1$, in this case there are $3$ outputs, which is reflected in Fig. 2.

\begin{table}[!t]
\caption{Training Dataset for Neural Network Model}
\label{table9}
\centering
\begin{tabular}{|c|c|}
\hline
        $R$  &                1,\,2,\,3,\,4,\,5,\,6,\,7,\,8 \\
\hline
        $\lambda_2$  &            1, \,2, \,3, \,4, \,5 \\
\hline
        $\mu_2$  &          8, \,10, \,12, \,14, \,16 \\
\hline
\end{tabular}

%\textbf{Table 9:} Training Dataset for Neural Network Model
\end{table}

\begin{figure}[!t]
\centering
\includegraphics[width=2.5in]{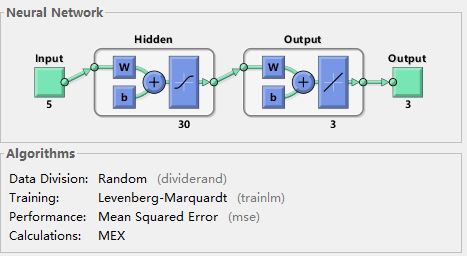}
\caption{Neural Network Model Created With Matlab}
\end{figure}

\begin{table}[!t]
\caption{Thresholds comparison from Neural Network to Actual Value}
\label{table10}
\centering
\begin {tabular} {|c|c|c|c|c|} 
\hline 
  & 	$R=1.3$ & 	2.3  & 	3.3 & 	4.3 	\\ 
\hline 
0 & 	(11, 11) & 	(13, 13) & 	(16, 16) & 	(18, 18) 	\\ 
\hline 
$n_1\downarrow$ & 	(9, 9) & 	(12, 12) & 	(14, 15) & 	(17, 17) 	\\ 
\hline 
2 & 	(8, 7) & 	(11, 11) & 	(13, 14) & 	(16, 16) 	\\ 
\hline 
  & 	$R=5.3$ & 	6.3  & 	7.3 & 	8.3 	\\ 
\hline 
0 & 	(20, 20) & 	(22, 22) & 	(23, 24) & 	(25, 25) 	\\ 
\hline 
$n_1\downarrow$ & 	(19, 18) & 	(21, 20) & 	(22, 22) & 	(24, 24) 	\\ 
\hline 
2 & 	(18, 18) & 	(19, 20) & 	(21, 21) & 	(23, 23) 	\\ 
\hline 
\end{tabular}

%\textbf{Table 10:} Thresholds comparison from Neural Network to Actual Value
\end{table}
After the model is trained, we can test the model with some other parameter settings which is different from those in the initial training dataset. If denote by $n_r$ as the real threshold values from Value Iteration method, $n_m$ as the threshold value through machine learning method, as listed in the Table IX, several parameter sets for $(R, \lambda_2=1, \mu_2=8)$ are chosen to compare the threshold pairs $(n_r, n_m)$ from actual computation and the neural network model. From TABLE X, it is observed that the machine learning model can do a good estimation of the thresholds.

%\newpage
\section{Conclusion and Discussion}

In order to reduce energy costs, cloud providers start to mix online and batch tasks in the same data center. However, it is always challengeable to efficiently optimize the assignment of the suitable resource to the users. To address this challenging problem, a novel model of a cloud data center is studied in this paper with cognitive capabilities for real-time (or online) flow compared to the batch tasks. Here, a DC can determine the cost of using resources. An online user or the user with batch tasks may decide whether or not to pay for getting the services. Particularly, the online service tasks have a higher priority over batch tasks and both types of tasks needs a certain number of virtual machines (VM). The objective here is to maximize the total discounted reward. The optimal policy to reach the objective for admitting task tasks is finally verified to be a state-related control limit policy.  After the optimal policy is determined in general, it is hard to identify the range of the threshold values which is particularly useful in reality. Therefore, we further consider the possible boundaries of the optimal value. A lower and an upper bound for such an optimal policy are then derived, respectively, for the estimation purpose. Finally, a comprehensive set of experiments on the various cases to validate this proposed solution is conducted. As a demonstration, the machine learning method is adopted to show how to obtain the optimal values by using a feed-forward neural network model. It is our observation that the major idea of this research can be extended to the research to maximize the expected average award or other related objective functions. The results offered in this paper will be expectedly utilized in various cloud data centers with different cognitive characteristics in an economically optimal strategy.

%\newpage

%\input{appendixWL.tex}
%\newpage
%\appendices
%\section
\appendix[Verification of Lemma 3]                                                                                 
We will now provide the Verification of Lemma 3. Similar as equation (\ref{Xcon1}), for $n_1 < N_1$, we have
%\small
\begin{eqnarray}
\label{XconB1A2}
& &  \big(\alpha + \beta_0(n_1+1, n_2+1)\big)
\Delta_{n_2} X(n_1+1, n_2)  \nonumber \\
&=&
  - \Delta_{n_2} f(n_1+1, n_2) \nonumber \\
%& & + \lambda_1 \Big[  X(n_1+1,n_2) - X(n_1,n_2) \Big] \nonumber\\
& & + \lambda_1  \Delta_{n_2}  v(\langle(n_1+1,n_2),A_1\rangle)  \nonumber\\
& & + \lambda_2  \Delta_{n_2}  v(\langle(n_1+1,n_2),A_2\rangle)  \nonumber \\
& &  + C_1(n_1+1)\mu_1  \Delta_{n_2} X(n_1,n_2)  \nonumber\\
& & + (C - C_1(n_1+1))\mu_2  \Delta_{n_2} X(n_1+1,n_2-1) . %\\
\end{eqnarray}
%\normalsize
%\begin{eqnarray}
%\label{XconB1A2}
%& &  \big(\alpha + \beta_0(n_1+1, n_2+1)\big)
%\Delta_{n_2} X(n_1+1, n_2)  \nonumber \\
%&=&
%  - \Delta_{n_2} f(n_1+1, n_2)  + \lambda_1  \Delta_{n_2}  v(\langle(n_1+1,n_2),A_1\rangle)   + \lambda_2  \Delta_{n_2}  v(\langle(n_1+1,n_2),A_2\rangle)  \nonumber \\
%& &  + C_1(n_1+1)\mu_1  \Delta_{n_2} X(n_1,n_2)   + (C - C_1(n_1+1))\mu_2  \Delta_{n_2} X(n_1+1,n_2-1) . %\\
%\end{eqnarray}
%For simplification of the mathematical expressions, in additional to the notations in equation (\ref{Delta1and2}), if for any function $g(n_1,n_2)$, we further introduce the notation
%\begin{eqnarray}
%\label{Delta1^2}
%& & \Delta^{(2)}_{n_1, n_2} g(n_1, n_2) \nonumber \\
%&=& \Delta_{n_2}g(n_1+1, n_2) - \Delta_{n_2}g(n_1, n_2). \\
%&=& \Delta_{n_1}g(n_1, n_2+1) - \Delta_{n_1}g(n_1, n_2).
%\end{eqnarray}

By a similar implementation on equation (\ref{Xcon1}) and equation (\ref{XconB1A2}), for $n_1 < N_1$, we have
%\small
\begin{eqnarray}
\label{XconcaveB1A}
& &  (\alpha + \beta_0(n_1+1, n_2+1)) \Delta^{(2)}_{n_1, n_2} X(n_1, n_2)    \nonumber \\
& = & (\alpha + \beta_0(n_1+1, n_2+1)) \Delta_{n_2} X(n_1+1, n_2) \nonumber \\
& & - (\alpha + \beta_0(n_1, n_2+1) + b\mu_1 -b\mu_2 ) \Delta_{n_2} X(n_1, n_2) \nonumber \\
&=&
- \Delta^{(2)}_{n_1, n_2} f(n_1, n_2) \nonumber \\
& & + \lambda_1 \Delta^{(2)}_{n_1, n_2} v(\langle(n_1,n_2),A_1\rangle)  \nonumber\\
& & + \lambda_2 \Delta^{(2)}_{n_1, n_2} v(\langle(n_1,n_2),A_2\rangle) \nonumber \\
& &  + C_1(n_1+1)\mu_1 \Delta_{n_2} X(n_1,n_2)  \nonumber \\
& & -  C_1(n_1)\Delta_{n_2} X(n_1-1,n_2)  \nonumber\\
& &  + C_2(n_1+1,n_2)\mu_2 \Delta_{n_2} X(n_1+1,n_2-1) \nonumber\\
& & - C_2(n_1,n_2)\Delta_{n_2} X(n_1,n_2-1)\nonumber\\
& & +b\mu_2 \Delta_{n_2} X(n_1, n_2) -b\mu_1 \Delta_{n_2} X(n_1, n_2)\nonumber\\
&=&
- \Delta^{(2)}_{n_1, n_2} f(n_1, n_2) \nonumber \\
& & + \lambda_1 \Delta^{(2)}_{n_1, n_2} v(\langle(n_1,n_2),A_1\rangle) \nonumber\\
& & + \lambda_2 \Delta^{(2)}_{n_1, n_2} v(\langle(n_1,n_2),A_2\rangle) \nonumber \\
& &  + C_1(n_1)\mu_1 \Delta^{(2)}_{n_1, n_2} X(n_1-1,n_2)  \nonumber\\
& &  + C_2(n_1+1,n_2)\mu_2 \Delta^{(2)}_{n_1, n_2} X(n_1,n_2-1) \nonumber\\
& &  +b\mu_2 \Delta^{(2)}_{n_2} X(n_1,n_2-1).%\\
\end{eqnarray}
%\normalsize
%\begin{eqnarray}
%\label{XconcaveB1A}
%& &  (\alpha + \beta_0(n_1+1, n_2+1)) \Delta^{(2)}_{n_1, n_2} X(n_1, n_2)    \nonumber \\
%& = & (\alpha + \beta_0(n_1+1, n_2+1)) \Delta_{n_2} X(n_1+1, n_2) \nonumber \\
%& & - (\alpha + \beta_0(n_1, n_2+1) + b\mu_1 -b\mu_2 ) \Delta_{n_2} X(n_1, n_2) \nonumber \\
%&=&
%- \Delta^{(2)}_{n_1, n_2} f(n_1, n_2)  + \lambda_1 \Delta^{(2)}_{n_1, n_2} v(\langle(n_1,n_2),A_1\rangle)   + \lambda_2 \Delta^{(2)}_{n_1, n_2} v(\langle(n_1,n_2),A_2\rangle) \nonumber \\
%& &  + C_1(n_1+1)\mu_1 \Delta_{n_2} X(n_1,n_2)  -  C_1(n_1)\Delta_{n_2} X(n_1-1,n_2)  \nonumber\\
%& &  + C_2(n_1+1,n_2)\mu_2 \Delta_{n_2} X(n_1+1,n_2-1)  - C_2(n_1,n_2)\Delta_{n_2} X(n_1,n_2-1)\nonumber\\
%& & +b\mu_2 \Delta_{n_2} X(n_1, n_2) -b\mu_1 \Delta_{n_2} X(n_1, n_2)\nonumber\\
%&=&
%- \Delta^{(2)}_{n_1, n_2} f(n_1, n_2) \nonumber \\
%& & + \lambda_1 \Delta^{(2)}_{n_1, n_2} v(\langle(n_1,n_2),A_1\rangle) \nonumber\\
%& & + \lambda_2 \Delta^{(2)}_{n_1, n_2} v(\langle(n_1,n_2),A_2\rangle) \nonumber \\
%& &  + C_1(n_1)\mu_1 \Delta^{(2)}_{n_1, n_2} X(n_1-1,n_2)  \nonumber\\
%& &  + C_2(n_1+1,n_2)\mu_2 \Delta^{(2)}_{n_1, n_2} X(n_1,n_2-1) \nonumber\\
%& &  +b\mu_2 \Delta^{(2)}_{n_2} X(n_1,n_2-1).%\\
%\end{eqnarray}

%\begin{eqnarray}
%\label{va1Ad}
%& & \Delta_{n_2} v(\langle(n_1,n_2),A_1\rangle) \nonumber \\
%&=&  \left\{\begin{array}{cl}
% \Delta_{n_2} X(\langle(n_1+1,n_2),A_1\rangle), & n_1 < N_1,\\
% %\Delta_{n_2} X(\langle(n_1+1,n_2),A_1\rangle) -r, & C - b < C_1 + C_2 \le C,\\
%  \Delta_{n_2} X(\langle(n_1,n_2),A_1\rangle), & n_1 = N_1.
%            \end{array}\right.
%\end{eqnarray}

We can now use Value Iteration Method with three steps to show that $\Delta_{n_2} X(n_1,n_2)$ is a nonincreasing function on $n_1$ for any given $n_2$ as below:

{\bf Step A-1:}
%Set $X^0(n_1,n_2) = 0$, by noting equations (\ref{va1}) and (\ref{va2}), we know $v^{(0)}(\langle(n_1,n_2),A_2\rangle)=R$ and
%\begin{eqnarray*}
%%\label{va1b}
% v^{(0)}(\langle(n_1,n_2),A_1\rangle)
%=  \left\{\begin{array}{cl}
% -br , & n_1 <N_1,\\
%  0, &  n_1 = N_1.
%            \end{array}\right.
%\end{eqnarray*}
%Substitute these  three results into equation (\ref{v01}), we will have
%\begin{eqnarray*}
%& & X^{(1)}(n_1,n_2)  \\
%&=&  \left\{\begin{array}{cl}
% \frac{-f(n_1, n_2) +(-br)+\lambda_2 R }{\alpha + c} , & n_1 <N_1,\\
%  \frac{-f(n_1, n_2) +0 +\lambda_2 R}{\alpha + c} , &  n_1 = N_1.
%            \end{array}\right.
%\end{eqnarray*}
%.
%Therefore, for any $n_2$, $X^{(1)}(n_1,n_2)$ is concave and nonincreasing on $n_1$
%Therefore, with $f(n_1,n_2)$ convex and nondecreasing on $n_1$,
%{\color{red} The condition we need is }
Similar as the analysis in Theorem 1, set $X^{(0)}(n_1,n_2) = 0$, we will have
$$\Delta_{n_2} X^{(1)} (n_1, n_2) = -\frac{\Delta_{n_2} f(n_1, n_2)}{\alpha + c}.$$
Since $\Delta_{n_2} f(n_1+1, n_2)\ge \Delta_{n_2} f(n_1, n_2))$, we will also have
\begin{eqnarray*}
& & \Delta_{n_2} X^{(1)}(n_1+1, n_2) - \Delta_{n_2} X^{(1)} (n_1, n_2))  \\
& = & \frac{\Delta_{n_2} f(n_1, n_2) - \Delta_{n_2} f(n_1+1, n_2))}{\alpha + c}  \\
&\le & 0.
\end{eqnarray*}%$X^{(1)}(n_1,n_2)$ is concave and nonincreasing on $n_2$.
%{\color{red} IT is a key point here, we are calculating $\Delta_{n_2} X^{(1)} (n_1, n_2))$ but not the difference along $n_1$ direction.}

{\bf Step A-2:}  By using the result in above Step 1 and the equation (\ref{vA_1}), we know that
$$
\Delta_{n_2} v^{(1)}(\langle(n_1+1,n_2),A_1\rangle) \le \Delta_{n_2} v^{(1)}(\langle(n_1,n_2),A_1\rangle).
$$
%$v^1(\langle n_1,n_2, A_1\rangle)$ is concave and non-increasing functions on $n_1$ for any $n_2$.
From the result and verification process in Theorem 1, with the cost function being convex and nondecreasing which means $\Delta_{n_2} f(n_1, n_2+1)\ge \Delta_{n_2} f(n_1, n_2)$, we know that $X^{(1)}(n_1,n_2)$ is concave and nonincreasing on $n_2$ for any $n_1$, which means
\begin{eqnarray*}
\Delta_{n_2} X^{(1)}(n_1,n_2) & \le & \Delta_{n_2} X^{(1)}(n_1,n_2-1), \\
\Delta_{n_2} X^{(1)}(n_1+1,n_2) & \le & \Delta_{n_2} X^{(1)}(n_1+1,n_2-1).
\end{eqnarray*}
    %and,
   By further applying the result in Lemma 2, let $X^{(1)}(n_1+1,i)$ be $h_1(i)$ and $X^{(1)}(n_1,i)$ be $h_2(i)$.
Since $\Delta_{n_2} X^{(1)}(n_1+1, n_2) \le \Delta_{n_2} X^{(1)} (n_1, n_2))$ for any $n_2$, by using equation (\ref{va2}), we will know that
   $$\Delta_{n_2} v^{(1)}(\langle(n_1+1,n_2),A_2\rangle) \le \Delta_{n_2} v^{(1)}(\langle(n_1,n_2),A_2\rangle). $$
By using the results in equation (\ref{XconcaveB1A}), we will have %s (\ref{XconB1A1}), (\ref{XconB1A2}) and
     $$\Delta_{n_2} X^{(2)}(n_1+1, n_2) - \Delta_{n_2} X^{(2)} (n_1, n_2))\leq 0. $$
%   These two inequalities justify that for any $n_1$,
%    $X^{(2)}(n_1,n_2)$ is  nonincreasing and concave on $n_1$.

%\item
{\bf Step A-3:} Finally, by noting the \emph{Theorem 11.3.2} of ~\cite{Puterman2005MDP} that the optimality equation has the unique solution, we know the value iteration $X^{(n)}(n_1,n_2)$ will uniquely converges. Therefore, as the iteration continues, with $n$ goes to $\infty$, we know that for any $n_1<N_1$,
$$\Delta_{n_2} X(n_1+1, n_2)\le \Delta_{n_2} X (n_1, n_2)),$$
always holds.

The verification of the Lemma 3 is now completed.

%\newpage
\bibliographystyle{IEEEtran}
\bibliography{IEEEabrv,nirefer}

\end{document}